\documentstyle[12pt]{article}

\begin{document}

\def\ft{1}
\def\ad{2}
\def\dh{3}
\def\bv{4}
\def\ab{5}
\def\jb{6}
\def\bg{7}
\def\at{8}
\def\ew{9}
\def\bk{10}
\def\hm{11}
\def\nt{12}
\def\ntt{13}

\begin{titlepage}

\begin{flushright}
hep-th/9305077
\end{flushright}

\vskip 0.6truecm

\begin{center}
{\large \bf ON QUANTUM INTEGRABILITY \\ \vskip 0.4cm AND THE
LEFSCHETZ NUMBER \\}
\end{center}

\vskip 1.0cm

\begin{center}
{\bf Antti J. Niemi$^{*}$ } \\
\vskip 0.4cm
{\it Department of Theoretical Physics, Uppsala University
\\ P.O. Box 108, S-75108, Uppsala, Sweden }
\vskip 0.5cm
and \\
\vskip 0.5cm
{\bf Kaupo Palo$^{**}$ } \\
\vskip 0.4cm
{\it Department of Theoretical Physics, Uppsala University
\\ P.O. Box 108, S-75108, Uppsala, Sweden \\}
\vskip 0.2cm
and
\vskip 0.2cm
{\it Institute of Physics, Estonian Academy of Sciences \\
142 Riia St., 202 400 Tartu, Estonia }

\end{center}

\vskip 2.3cm

\rm
\noindent
Certain phase space path integrals can be evaluated exactly
using equivariant cohomology and localization in the
canonical loop space.  Here we extend this to a general
class of models. We consider hamiltonians which are {\it a
priori} arbitrary functions of the Cartan subalgebra
generators of a Lie group which is defined on the phase
space.  We evaluate the corresponding path integral and find
that it is closely related to the infinitesimal Lefschetz
number of a Dirac operator on the phase space. Our results
indicate that equivariant characteristic classes could
provide a natural geometric framework for understanding
quantum integrability.

\vfill

\begin{flushleft}
\rule{5.1 in}{.007 in}\\
$^{*}$ {\small E-mail: NIEMI@TETHIS.TEORFYS.UU.SE \\}
$^{**}$ {\small E-mail: PALO@RHEA.TEORFYS.UU.SE \\ }
\end{flushleft}

\end{titlepage}

\vfill\eject

\baselineskip 0.65cm

There are several different approaches to quantum
integrability [\ft,\ad].  Here we shall investigate a
geometric approach which is based on phase space path
integrals and loop space equivariant cohomology [\dh-\ntt].
We show that the formalism is applicable to a general class
of integrable models with hamiltonians that are functionals
of canonical Lie algebra generators. Our result extends
those derived previously in [\nt,\ntt]. In particular, we
find that for the class of models we consider the path
integral can be evaluated exactly using localization
techniques [\dh-\ntt]. The final result is an ordinary
integral over the phase space, and can be represented as a
functional of an equivariant characteristic class [\bg].  We
also find that the path integral can be related to the path
integral representation of a Dirac operator which is defined
on the phase space of the theory. In particular, the path
integral of our integrable models is closely related to the
infinitesimal Lefschetz number which is associated to this
Dirac operator. Our results suggest that equivariant
cohomology and equivariant characteristic classes seem to
provide a natural framework for a geometric formulation of
quantum integrability.

We consider a hamiltonian $H$ on a phase space $\Gamma$ with
$2n$ coordinates $z^{a}$ and Poisson bracket
$$
\{ z^{a} , z^{b} \}_{\omega} ~=~ \omega^{ab}(z)
\eqno (1)
$$
Here the inverse matrix $\omega_{ab}$ determines the
symplectic two-form on the phase space.  We shall assume
that on $\Gamma$ there is a canonical realization of a Lie
algebra by functions $G_{u}(z)$,
$$
\{ G_{u} , G_{v} \}_{\omega} ~=~ \omega^{ab} \partial_{a}
G_{u} \partial_{b} G_{v} ~=~ {f_{uv}}^{w} G_{w}
\eqno (2)
$$
and we shall assume that the hamiltonian $H$ is a functional
of these generators,
$$
H (z) ~=~ H [G_{u}(z)]
\eqno (3)
$$

An integrable hamiltonian $H$ can be naturally viewed as a
functional of the Cartan subalgebra of some Lie algebra (2).
Indeed, for an integrable hamiltonian $H$ we can find
canonically conjugated action ($I_{a}$) and angle
($\theta^{a}$) variables
$$
\{ I_{a} , \theta^{b} \}_{\omega} ~=~ \delta_{a}^{b}
\eqno (4)
$$
such that the hamiltonian $H$ is a functional of the action
variables $I_{a}$ only. These action variables label
invariant torii on the phase space and the motion of $H(I)$
is constrained on these torii.  Since the action variables
are in involution,
$$
\{ I_{a} , I_{b} \}_{\omega} ~=~ 0
\eqno (5)
$$
we can interpret them as generators in the Cartan subalgebra
of some Lie algebra (2).

As an example consider $S^{2}$, which is the co-adjoint
orbit of SU(2).  With $z$ and $\bar z$ the coordinates on
the Riemann sphere $S^{2}$ the Poisson bracket is
$$
\{ z , \bar z \} ~=~ \frac{1}{2iJ} (1+ |z|^{2})^{2}
\eqno (6)
$$
This corresponds to the canonical spin-$J$ representation of
SU(2) with generators
$$
S_{0} ~=~ - J { 1- |z|^{2} \over 1+|z|^{2} } ~~~~~~~~ S_{+}
{}~=~ 2J {
\bar z \over 1+ |z|^{2} } ~~~~~~~~S_{-} ~=~ 2J { z \over 1 +
|z|^{2} }
\eqno (7)
$$
Our hamiltonian $H(G_{u})$ is then a generic functional of
all generators (7), while an integrable hamiltonian is
obtained {\it e.g.} by considering functions of $S_{0}$
only.

We are interested in the evaluation of the canonical phase
space path integral
$$
Z ~=~ \int [dz^{a}] \sqrt{ || \omega_{ab}|| } exp\{ i
\int\limits_{0}^{\beta} \vartheta_{a} {\dot z}^{a} - H(G_{u}) \}
\eqno (8)
$$
for a generic classically integrable hamiltonian $H(G_{u})$.
Here $\vartheta_{a}$ are components of the symplectic
one-form,
$$
\omega_{ab} ~=~ \partial_{a} \vartheta_{b} -
\partial_{b}\vartheta_{a}
\eqno (9)
$$

In order to evaluate (8) we introduce arbitrary functions
$j^{u}(t)$ that are independent of the phase space
coordinates, and use familiar identities to write (8) as
$$
Z ~=~ \exp \{ - i \int\limits_{0}^{\beta} H \left[
\frac{1}{i}\frac{\delta}{\delta j^{u}(t)} \right] \}
\int [dz^{a}] \sqrt{ || \omega_{ab}|| }
exp\{ i \int\limits_{0}^{\beta} \vartheta_{a} {\dot z}^{a} -
j^{u}G_{u}
\}_{|_{j=0}}
\eqno (10)
$$
Consequently it is sufficient to evaluate the following path
integral
$$
Z_{o} (j) ~=~ \int [dz^{a}] \sqrt{ || \omega_{ab}|| } exp\{
i
\int\limits_{0}^{\beta} \vartheta_{a} {\dot z}^{a} - j^{u}G_{u}
\}
\eqno (11)
$$
with arbitrary functions $j^{u}(t)$.

For $j^{u} $ that are $t$-independent, (11) yields the Weyl
character of the Lie group in the given representation. For
arbitrary $t$-dependent functions $j^{u}(t)$ we consider an
infinitesimal variation of the action in (11),
$$
\delta(\vartheta_{a} {\dot z}^{a} - j^{u} G_{u} ) ~=~
\delta z^{a} (\omega_{ab} {\dot z}^{b} - j^{u} \partial_{a}
G_{u})
\eqno (12)
$$
We select
$$
\delta z^{a} ~=~ \epsilon^{u} \{ G_{u} , z^{a} \}_{\omega}
\eqno (13)
$$
where the $\epsilon^{u}$ are some infinitesimal
$z^{a}$-independent parameters. This corresponds to the
infinitesimal limit of the canonical transformation
$$
z^{a} ~\to~ e^{-\epsilon^{u} G_{u} } z^{a} e^{\epsilon^{u}
G_{u} } ~=~ z^{a} + \epsilon^{u} \{ z^{a} , G_{u} \} +
\frac{1}{2}
\epsilon^{u} \epsilon^{v} \{ \{ z^{a} , G_{u} \} , G_{v} \} +
...
\eqno (14)
$$
and yields
$$
\delta ( \vartheta_{a} {\dot z}^{a} - j^{u} G_{u} )
{}~=~ - {\dot \epsilon}^{u} G_{u} - {f_{uv}}^{w} \epsilon^{u}
j^{v} G_{w}
\eqno (15)
$$
Since the Liouville measure in (11) is invariant under
canonical transformations we then conclude that the only
effect of (13) and (15) in (11) is that the external sources
$j^{u}(t)$ are shifted according to
$$
j^{u} ~\to~ j^{u} + {\dot \epsilon}^{u} + {f_{vw}}^{u}
\epsilon^{v} j^{w}
\eqno (16)
$$
Notice that (16) is analogous to the nonabelian gauge
transformation in a Yang-Mills theory. Indeed, if we relate
the generators $G_{u}$ to the Gauss law generators $G_{u}
\sim D_{i}^{ab} E_{i}^{b}$ in a nonabelian gauge theory and
identify $j^{u}(t)$ with the time component of the gauge
field $j^{u} \sim A_{o}^{a}$, we observe that (16) has the
functional form of a time dependent gauge transformation.
On the basis of this analogy we then conclude that we can
indeed expect the measure in (11) to be invariant under
(13), and if it for some reason fails to be invariant we
expect that the ensuing nontrivial Jacobian can be related
to conventional nonabelian gauge anomalies.

As a consequence the path integral can only depend on gauge
invariant functionals of $j^{\mu}$ {\it i.e.} the temporal
Wilson loop. For the Cartan subalgebra $T_{u}$ of $G_{u}$
$$
\{ T_{u} , T_{v} \} ~=~ 0
\eqno (17)
$$
corresponding to integrable models, we can then further
simplify (16) to
$$
j^{u} ~\to~ j^{u} + {\dot \epsilon}^{u}
\eqno (18)
$$
and if we divide $j^{u}(t)$ into its constant part and its
$t$-dependent part
$$
j^{u}(t) ~=~ j^{u}_{o} + j^{u}_{t}
\eqno (19)
$$
where $j^{u}_{o}$ denotes the constant part and $j^{u}_{t}$
the $t$-dependent part {\it e.g.} in a Fourier
decomposition, we can gauge transform the path integral (11)
to
$$
Z_{o}(j) ~=~ \int [d z^{a}] \sqrt{ det || \omega_{ab} || }
\exp \{ i \int \vartheta_{a} {\dot z}^{a} - j^{u}_{o} T_{u}  \}
\eqno (20)
$$
{\it i.e.} it depends {\it only} on the constant modes
$j^{u}_{o}$.  Since (10) is evaluated at $j^{u}(t) = 0$, we
conclude that in the abelian case the original path integral
(8) simplifies to
$$
Z ~=~ \exp\{ - i \beta H \left[ \frac{1}{i} {\partial \over
\partial j^{u}_{o} } \right] \} \int [dz^{a}] \sqrt{ det ||
\omega_{ab} || }
\exp\{ i \int \vartheta_{a} {\dot z}^{a} - j^{u}_{o} T_{u}
\}_{|_{j_{o}=0}}
\eqno (21)
$$
Consequently it is sufficient to evaluate the path integral
$$
Z(j_{o}) ~=~ \int [dz^{a}] [dc^{a}]
\exp\{ i \int \vartheta_{a} {\dot z}^{a} - T +
\frac{1}{2} c^{a}\omega_{ab} c^{b} \}
\eqno (22)
$$
where we have introduced anticommuting variables $c^{a}$ to
exponentiate the determinant of the symplectic two-form and
$$
T = j^{u}_{o} T_{u}
\eqno (23)
$$
is an element of the Cartan subalgebra {\it i.e.} a U(1)
generator of the Lie algebra,

The path integral in (22) yields the Weyl character of $T$
and has been evaluated {\it e.g.} in [\nt,\ntt] using
localization methods.  Here we need a few relevant steps: We
interpret (22) as an integral in the loop space of the phase
space, parametrized by the time evolution $z^{a} \to
z^{a}(t)$. We identify $c^{a}(t)$ as a basis of loop space
one-forms and interpret the bosonic part of the action in
(22) as a loop space hamiltonian functional. The
corresponding loop space hamiltonian vector field then has
components
$$
{\cal X}_{S}^{a} ~=~ \dot z^{a} - \omega^{ab}
\partial_{b} T
\eqno (24)
$$
We introduce the loop space exterior derivative (in the
following time integrals are implicit)
$$
d ~=~ c^{a} \partial_{a}
\eqno (25)
$$
and a basis $i_{a}$ of loop space contractions dual to the
basis of one-forms $c^{a}(t)$,
$$
i_{a}(t)c^{b}(t') ~=~ \delta_{a}^{b}(t-t')
\eqno (26)
$$
We then define the following equivariant exterior derivative
on the loop space,
$$
d_{S} ~=~ d+ i_{S} ~=~ c^{a} \partial_{a} + {\cal X}_{S}^{a}
i_{a}
\eqno (27)
$$
The corresponding Lie-derivative along the hamiltonian
vector field (24) is
$$
{\cal L}_{S} ~=~ d i_{S} + i_{S} d
\eqno (28)
$$
On the subspace of the exterior algebra where (28) vanishes,
(27) is then nilpotent.

The action in (22) is closed with respect to (27),
$$
d_{S} (\vartheta_{a} {\dot z}^{a} - T + \frac{1}{2} c^{a}
\omega_{ab} c^{b} )
\eqno (29)
$$
In [\bk] it has been shown, that if we add to the action a
$d_{S}$ exact term,
$$
S ~\to~ \int \vartheta_{a} {\dot z}^{a} - T + \frac{1}{2}
c^{a}
\omega_{ab} c^{b} + d_{S} \Psi
\eqno (30)
$$
where $\Psi$ is in the nilpotent subspace of (27),
$$
{\cal L}_{S} \Psi ~=~ 0
\eqno (31)
$$
the corresponding path integral
$$
Z_{\Psi} ~=~ \int [dz^{a}] [dc^{a}] exp \{ i \int
\vartheta_{a} {\dot z}^{a} - T + \frac{1}{2}
c^{a}\omega_{ab} c^{b} + d_{S}
\Psi \}
\eqno (32)
$$
is independent of $\Psi$ and coincides with (22). By
selecting a suitable $\Psi$ the path integral can then be
evaluated by the localization method.

In order to construct the appropriate $\Psi$ we observe that
the phase space $\Gamma$ admits a metric tensor $g_{ab}$
which is Lie-derived by the U(1) generator $T$,
$$
{\cal L}_{T} g ~=~ 0
\eqno (33)
$$
For this, it is sufficient to take any metric tensor on
$\Gamma$ and average it over the group. This yields a metric
tensor that satisfies (33).  If we then select
$$
\Psi ~=~ \frac{1}{2} g_{ab} {\dot z}^{a} c^{b}
\eqno (34)
$$
the condition (31) is satisfied and the corresponding action
(32) is
$$
S ~=~ \int \frac{1}{2} g_{ab} \dot z^{a}
\dot z^{b}
+ (\vartheta_{a} - \frac{1}{2} g_{ab}{\cal X}_{T}^{b})
\dot z^{a} - T
- \frac{1}{2} c^{a} ( g_{ab} \partial_{t} +
\dot z^{c}g_{bd}\Gamma^{d}_{ac} ) c^{b}
+ \frac{1}{2} c^{a} \omega_{ab} c^{b}
\eqno (35)
$$
Here $\Gamma^{d}_{ac}$ is the Christoffel symbol for the
metric tensor $g_{ab}$. We scale $g_{ab} \to
\lambda g_{ab}$.  According to the general argument [\bk] the
path
integral is independent of $\lambda$.  Hence we can evaluate
(32) in the $\lambda
\to \infty$ and the result is [\nt,\ntt]
$$
Z_{o} ~=~ \int dz_{o} dc_{o} \exp\{ -i \beta T + i
\frac{\beta}{2} c^{a}_{o} \omega_{ab} c^{b}_{o} \}
{ 1 \over \sqrt{ det [ {\delta^{a}}_{b} \partial_{t} -
\frac{1}{2} ({\Omega^{a}}_{b} + {R^{a}}_{b}) ] } }
\eqno (36)
$$
where
$$
{\Omega^{a}}_{b} ~=~ \frac{1}{2} g^{ac} [ \partial_{b}
(g_{cd} {\cal X}^{d}_{T}) - \partial_{c}(g_{bd} {\cal
X}^{d}_{T}) ]
\eqno (37)
$$
is the Riemannian momentum map of $T$ [\bg] and ${R^{a}}_{b}
= {R^{a}}_{bcd}c_{o}^{c} c_{o}^{d}$ is the Riemannian
curvature two-form.  If we evaluate the determinant
using{\footnotemark\footnotetext{ Here we use a
regularization which is symmetric in the negative and
positive eigenvalues of the differential operator in (36)}
{\it e.g.} $\zeta$-function regularization we get
$$
Z_{o} ~=~ \int dz_{o} dc_{o} exp\{ -i \beta T + i
\frac{\beta}{2} c^{a}_{o} \omega_{ab} c^{b}_{o} \}
\sqrt{ det\left[ { \frac{\beta}{2} ({\Omega^{a}}_{b} +
{R^{a}}_{b}
) \over sinh[ \frac{\beta}{2} ( {\Omega^{a}}_{b} +
{R^{a}}_{b} ) ] }
\right] }
\eqno (38)
$$
Here
$$
Ch\left[ \frac{\beta}{2} ( T-\omega) \right] ~=~ \exp \{ - i
\beta T + i
\frac{\beta}{2} c_{o}^{a} \omega_{ab} c_{o}^{b} \}
\eqno (39)
$$
is the equivariant Chern character and
$$
{\hat A}\left[ \frac{\beta}{2} ( {\Omega^{a}}_{b} +
{R^{a}}_{b} )
\right] ~=~ \sqrt{ det\left[ { \frac{\beta}{2} (
{\Omega^{a}}_{b}
+ {R^{a}}_{b} ) \over sinh[ \frac{\beta}{2} (
{\Omega^{a}}_{b} + {R^{a}}_{b} ) ] } \right] }
\eqno (40)
$$
is the equivariant ${\hat A}$-genus. Both (39) and (40) are
equivariantly closed,
$$
(d + i_{T})Ch\left[ \frac{\beta}{2} ( T - \omega) \right]
{}~=~ ( d + i_{T}) {\hat A} \left[ \frac{\beta}{2} (
{\Omega^{a}}_{b} + {R^{a}}_{b} ) \right] ~=~ 0
\eqno (41)
$$
If we substitute these into (21) we then obtain our
integration formula for the path integral (8), (21)
$$
Z~=~exp \{- i \beta \cdot H \left[ \frac{1}{i} {\partial
\over
\partial j^{u}_{o} } \right] \} \cdot ~Ch\left[ \frac{\beta}{2}
(j^{u}_{o} T_{u} - \omega) \right] {\hat A} \left[
\frac{\beta}{2} ( j^{u}_{o} {{[\Omega_{T_{u}}]}^{a}}_{b} +
{R^{a}}_{b} )
\right]_{|_{j=0}}
\eqno (42)
$$
{\it i.e.} the path integral reduces to a derivative
expansion of equivariant characteristic classes.

We shall now argue, that (42) is closely related to the
(infinitesimal) Lefschetz number of a Dirac operator which
is defined on the phase space $\Gamma$. For this we first
consider the following Dirac operator on an even dimensional
Riemannian manifold ${\cal M}$
$$
\gamma^{\mu} D_{\mu} ~=~ \left( \matrix{ 0 & D_{+}
\cr D_{-} & 0 \cr } \right)
{}~=~ \gamma^{\mu} ( \partial_{\mu} + A_{\mu} +
\frac{1}{8} \omega_{\mu ij} [ \gamma^{i} , \gamma^{j} ] )
\eqno (43)
$$
Here $\gamma^{i} = e_{\mu}^{i}\gamma^{\mu}$ are local Dirac
matrices, $e_{\mu}^{i}$ are components of the vielbein and
$\omega_{\mu ij}$ are components of the spin connection. We
shall assume that the manifod ${\cal M}$ admits an operator
realization of the Lie algebra (2),
$$
[{\cal G}_{u} , {\cal G}_{v} ] ~=~ {f_{uv}}^{w} {\cal G}_{w}
\eqno (44)
$$
such that the operators ${\cal G}_{u}$ commute with the
Dirac operator (43)
$$
[ {\cal G}_{u} , \gamma^{\mu} D_{\mu}] ~=~ 0
\eqno (45)
$$
The eigenstates of $\gamma^{\mu} D_{\mu}$ that correspond to
a fixed eigenvalue $\lambda$ then define a representation of
the Lie algebra ${\cal G}_{u}$. We shall be particularly
interested in the representation which is defined by the
zeromodes $\lambda=0$ of $\gamma^{\mu} D_{\mu}$.  For this
we introduce some arbitrary parameters $\alpha_{u}$ and
construct the linear combination $\alpha
\cdot {\cal G}$ which is an element of the Cartan subalgebra of
(44)
in a suitable basis. We then introduce the following
Lefschetz number,
$$
L (\alpha \cdot {\cal G}) ~=~ \lim_{\beta\to \infty} Tr
\{e^{i\alpha
\cdot {\cal G} } ( e^{- {\beta} D_{+}D_{-}} - e^{- {\beta}
D_{-}D_{+} }
) \}
\eqno (46)
$$
One can show that the trace on the {\it r.h.s.} of (46) is
independent of $\beta$ and provided either $D_{+}D_{-}$ or
$D_{-}D_{+}$ has no zero modes it coincides with the
character of the (reducible or irreducible) representation
of ${\cal G}_{u}$ determined by the zeromodes of (43).

We wish to relate the evaluation of the Lefschetz number
(46) to the evaluation of the path integral (8). For this we
need the appropriate canonical realization of the Dirac
operator $\gamma^{\mu} D_{\mu}$ and the Lie algebra
generators ${\cal G}_{u}$: We interpret the coordinates
$z^{\mu}$ on ${\cal M}$ as canonical coordinates on a phase
space with conjugate momenta $p_{\mu}$ and Poisson brackets
$$
\{ p_{\mu} , z^{\nu} \} ~=~ \delta_{\mu}^{\nu}
\eqno (47)
$$
We also realize the local $\gamma$-matrix algebra
canonically using anticommuting variables $\psi^{i}$ with
graded Poisson brackets
$$
\{ \psi^{i} , \psi^{j} \} ~=~ \eta^{ij}
\eqno (48)
$$
where $\eta^{ij}$ is the local flat metric. The
corresponding ${\cal M}$-dependent variables $\psi^{\mu} =
E_{i}^{\mu}
\psi^{i}$ have the brackets
$$
\{ \psi^{\mu} , \psi^{\nu} \} ~=~ g^{\mu\nu}
\eqno (49)
$$
where $g_{\mu\nu}$ is the metric tensor on ${\cal M}$. The
appropriate canonical realization of the Dirac operator (43)
is then
$$
{\cal Q} ~=~ \psi^{\mu} {\cal P}_{\mu} ~=~ \psi^{\mu} (
p_{\mu} +
\frac{1}{4} \omega_{\mu ij} \psi^{i} \psi^{j} + A_{\mu} )
\eqno (50)
$$
and
$$
\{ {\cal Q} , {\cal Q} \} ~=~ {\cal H} ~=~
g_{\mu\nu} {\cal P}^{\mu} {\cal P}^{\nu} + \frac{1}{2}
\psi^{\mu}\psi^{\nu}
F_{\mu\nu} ~=
$$
$$
{}~=~ g^{\mu\nu}(p_{\mu} + \frac{1}{4} \omega_{\mu ij}
\psi^{i} \psi^{j} + A_{\mu}) (p_{\nu} + \frac{1}{4} \omega_{\mu
kl}
\psi^{k} \psi^{l} + A_{\nu} ) + \frac{1}{2} \psi^{\mu}\psi^{\nu}
F_{\mu\nu}
\eqno (51)
$$
where we have used identities of the Riemann tensor.

We first consider the path integral for the Atiyah-Singer
index of (43), {\it i.e.} we set $\alpha = 0$ in (46). This
path integral is (see {\it e.g.} [\hm])
$$
\int
[dz^{\mu}] [dp_{\mu}][d\psi^{\mu}] exp\{ i \int p_{\mu}
{\dot z}^{\mu} -
\frac{1}{2} \psi_{\mu} {\dot \psi}^{\mu} - g_{\mu \nu} {\cal
P}^{\mu}{\cal P}^{\nu} - \frac{1}{2} \psi^{\mu} \psi^{\nu}
F_{\mu\nu}
\}
\eqno (52)
$$
We change variables
$$
p_{\mu} ~\to~ p_{\mu} - A_{\mu} - \omega_{\mu ij} \psi^{i}
\psi^{j}
\eqno (53)
$$
This gives
$$
S ~=~ \int p_{\mu} {\dot z}^{\mu} - A_{\mu} {\dot z}^{\mu} -
\frac{1}{2} g_{ \mu\nu } p^{\mu} p^{\nu} - \frac{1}{2}
\psi^{\mu}
(g_{\mu\nu}
\partial_{t} + {\dot z}^{\rho} g_{\mu\sigma}
\Gamma^{\sigma}_{\rho \nu} ) \psi^{\nu}
- \frac{1}{2} \psi^{\mu} \omega_{\mu\nu}\psi^{\nu}
\eqno (54)
$$
and eliminating $p_{\mu}$ we get
$$
S ~=~ \int \frac{1}{2} g_{\mu\nu} {\dot z}^{\mu} {\dot
z}^{\nu} - A_{\mu} {\dot z}^{\mu} - \frac{1}{2} \psi^{\mu}
(g_{\mu\nu}
\partial_{t} + {\dot z}^{\rho} g_{\mu\sigma}
\Gamma^{\sigma}_{\rho\nu} ) \psi^{\nu}
- \frac{1}{2} \psi^{\mu} F_{\mu\nu}\psi^{\nu}
\eqno (55)
$$
With the identification $A_{\mu} \sim - \vartheta_{a}$ and
$F_{\mu\nu}
\sim - \omega_{ab}$ this  coincides with the action in (35) with
$T=0$.

We shall now proceed to demonstrate that if $\alpha \not= 0$
in (46), its path integral representation coincides with
(35) for an appropriate $T\not= 0$. For this we need an
operator realization of the Lie algebra (44) in the extended
phase space. If we define
$$
{\cal G}_{u} ~=~ {\cal X}_{u}^{\mu} {\partial \over \partial
z^{\mu} } - {{\cal X}_{u}^{\mu}},_{\nu} p^{\nu} {\partial
\over
\partial p^{\mu} } + {{\cal X}_{u}^{\mu}},_{\nu} \psi^{\nu}
{\partial \over \partial
\psi^{\mu} }
\eqno (56)
$$
where ${\cal X}_{u}^{\mu}(z)$ are components of the
hamiltonian vector fields corresponding to the functions
$G_{u}(z)$ in (2), we find that (56) satisfy the Lie algebra
(44).

{}From (54) we extract the following symplectic one-form on
the extended phase space,
$$
\Theta ~=~ ( p_{\mu} + \vartheta_{\mu} - \frac{1}{2}\psi^{\nu}
g_{\nu\sigma} {\Gamma}^{\sigma}_{\mu \rho} \psi^{\rho})
dz^{\mu} -
\frac{1}{2}
\psi^{\mu} g_{\mu\nu} d\psi^{\nu}
\eqno (57)
$$
We shall assume that in the original phase space the
symplectic one-form $\vartheta_{\mu}(z)dz^{\mu}$ is Lie
derived by the vector fields ${\cal X}^{a}_{u}$,
$$
{\cal L}_{G_{u}} \vartheta ~=~ 0
\eqno (58)
$$
As a consequence we find that in the extended phase space
(56) Lie derives (57),
$$
{\cal L}_{{\cal G}_{u}} \Theta ~=~ 0
\eqno (59)
$$
Integrating this we then obtain the momentum map
$$
i_{ {\cal G}_{u} } \Theta ~=~ {\cal T}_{u} ~=~ {\cal
X}^{\mu}_{u} p_{\mu} + \vartheta_{\mu} {\cal X}^{\mu}_{u} -
\frac{1}{2}
\psi^{\mu}
\Omega^{u}_{\mu \nu} \psi^{\nu}
\eqno (60)
$$
where
$$
\Omega^{u}_{\mu\nu} ~=~ \frac{1}{2} [ \partial_{\mu}
( g_{\nu\sigma} {\cal X}_{u}^{\sigma} ) - \partial_{\nu} (
g_{\mu\sigma} {\cal X}_{u}^{\sigma} ) ]
\eqno (61)
$$
is the Riemannian momentum map for the generator $G_{u}$. In
the extended phase space equipped with the symplectic
structure defined by (57) the momentum maps (60) determine a
Poisson bracket realization of (2),
$$
\{ {\cal T}_{u} , {\cal T}_{v} \} ~=~ {f_{uv}}^{w} {\cal T}_{w}
\eqno (62)
$$
Furthermore, since
$$
\{ {\cal T}_{u} , {\cal Q} \} ~=~ 0
\eqno (63)
$$
we conclude that the path integral for the action
$$
S ~=~ \int p_{\mu} {\dot z}^{\mu} + \vartheta_{\mu} \dot
z^{\mu} -
\frac{1}{2} \psi^{\mu}(  g_{\mu\nu} \partial_{t} +
g_{\mu\sigma} \dot z^{\rho} {\Gamma}^{\sigma}_{\rho \nu})
\psi^{\nu}  +  \frac{1}{2}
\omega_{\mu\nu} \psi^{\mu}\psi^{\nu} - \frac{1}{2}g_{\mu\nu}
p^{\mu}
p^{\nu} - \tilde\alpha \cdot {\cal T}
\eqno (64)
$$
where $\tilde \alpha = \beta^{-1} \alpha$, yields the
Lefschetz number (46).  Eliminating $p_{\mu}$ we get further
$$
S ~=~ \int \frac{1}{2} g_{\mu\nu} {\dot z}^{\mu} {\dot
z}^{\nu} + {\dot z}^{\mu} [\vartheta_{\mu} + g_{\mu\nu}
(\tilde\alpha \cdot {\cal X}^{\mu}) ] + \frac{1}{2}
g_{\mu\nu} (\tilde\alpha \cdot {\cal X}^{\mu} )
(\tilde\alpha \cdot {\cal X}^{\nu} ) + (\tilde\alpha \cdot
{\cal X}^{\mu}) \vartheta_{\mu} -
$$
$$
- \frac{1}{2} \psi^{\mu}(g_{\mu\nu} \partial_{t} + {\dot
z}^{\rho}g_{\mu\sigma} \Gamma^{\sigma}_{\rho \nu})
\psi^{\nu} +
\frac{1}{2} \psi^{\mu} \omega_{\mu\nu} \psi^{\nu} +\frac{1}{2}
\psi^{\mu} (\tilde\alpha \cdot \Omega_{\mu\nu} )\psi^{\nu}
\eqno (65)
$$
If we select $\vartheta_{\mu}$ so that $T$ is obtained as
the momentum map
$$
T ~=~ - (\tilde \alpha {\cal X}^{\mu} ) \vartheta_{\mu}
\eqno (66)
$$
we then find
$$
S ~=~ \int \frac{1}{2} g_{\mu\nu} {\dot z}^{\mu} {\dot
z}^{\nu} + {\dot z}^{\mu} (\vartheta_{\mu} + g_{\mu\nu}
{\cal X}^{\mu}_{T} ) +
\frac{1}{2} g_{\mu\nu} {\cal X}^{\mu}_{T} {\cal X}^{\nu}_{T} - T
-
$$
$$
- \frac{1}{2} \psi^{\mu}(g_{\mu\nu} \partial_{t} + {\dot
z}^{\rho}g_{\mu\sigma} \Gamma^{\sigma}_{\rho \nu})
\psi^{\nu} +
\frac{1}{2} \psi^{\mu} \omega_{\mu\nu} \psi^{\nu} + \frac{1}{2}
\psi^{\mu}  \Omega^{T}_{\mu\nu} \psi^{\nu}
\eqno (67)
$$
This action has the following interpretation: From (55) we
conclude that the action
$$
S_{1} ~=~ \int \frac{1}{2} g_{\mu\nu} {\dot z}^{\mu} {\dot
z}^{\nu} -
\frac{1}{2} \psi^{\mu}(g_{\mu\nu} \partial_{t} +
{\dot z}^{\rho}g_{\mu\sigma} \Gamma^{\sigma}_{\rho \nu})
\psi^{\nu}
\eqno (68)
$$
can be interpreted as a Dirac operator on the Riemannian
manifold.  Similarly, the path integral for
$$
S_{2} ~=~ \int \vartheta_{\mu} {\dot z}^{\mu} - T +
\frac{1}{2}
\psi^{\mu} \omega_{\mu\nu} \psi^{\nu}
\eqno (69)
$$
yields the Weyl character of $T$. Finally, the action
$$
S_{3} ~=~ \int g_{\mu\nu} {\cal X}^{\mu}_{T} {\dot z}^{\mu}
-
\frac{1}{2} g_{\mu\nu} {\cal X}^{\mu}_{T} {\cal X}^{\nu}_{T} +
\frac{1}{2}
\psi^{\mu}  \Omega^{T}_{\mu\nu} \psi^{\nu}
\eqno (70)
$$
defines a bi-hamiltonian pair of $S_{2}$ in the sense that
in the place of the symplectic structure (9) we have
$$
\partial_{\mu} ( \frac{1}{2} g_{\nu\rho} {\cal X}^{\rho}_{T} ) -
\partial_{\nu} ( \frac{1}{2} g_{\mu\rho} {\cal X}^{\rho}_{T} )
{}~=~ \Omega^{T}_{\mu\nu}
\eqno (71)
$$
and in the place of the hamiltonian $T$ we have
$$
{\cal T} ~=~ \frac{1}{2} \psi^{\mu} \Omega^{T}_{\mu\nu}
\psi^{\nu}
\eqno (72)
$$
and the classical equations of motion for the pairs
$(T,\omega)$ and ${\cal T},
\Omega)$ coincide,
$$
{\dot z}^{\mu} ~=~ \{ T , z^{\mu} \}_{\omega} ~=~ \{ {\cal
T} , z^{\mu}
\}_{\Omega}
\eqno (73)
$$
The existence of this bi-hamiltonian structure is of course
consistent with the classical integrability of the
hamiltonian system $(T,
\omega)$.

In order to relate the actions (35) and (67) we introduce
$$
{\tilde \Psi} ~=~ \frac{1}{2} g_{\mu\nu} {\cal X}^{\mu}_{T}
\psi^{\nu}
\eqno (74)
$$
and add the corresponding $d_{S}$-exact term to the action
(35),
$$
S ~ \to ~ S + d_{S} {\tilde \Psi}
\eqno (75)
$$
This action then coincides with (67), and provided either
$D_{+}D_{-}$ or $D_{-}D_{+}$ in (46) has no zeromodes, the
corresponding path integrals must also coincide (possibly
modulo an overall sign).  Furthermore, from (22) and (32) we
conclude that the path integral for (67) differs from the
path integral for the Weyl character only by a $d_{S}$ exact
term. Consequently the infinitesimal Lefschetz number (46)
of our Dirac operator coincides with the Weyl character of
$T$, provided either $D_{+}D_{-}$ or $D_{-}D_{+}$ in (46)
has no zeromodes and the boundary conditions for the path
integrals have bee properly selected .

\vskip 0.4cm

In conclusion, we have generalized the localization
evaluation of path integrals discussed in [\nt,\ntt] to a
general class of hamiltonians which are constructed from Lie
algebra generators.  We have also established that the path
integral for these hamiltonians evaluates to equivariant
characteristic classes, and is also closely related to the
Lefschetz number of a Dirac operator. Our results indicate
that equivariant cohomology could provide a natural
geometric framework for understanding quantum integrability.

\vfill\eject

We thank A. Alekseev, L. Faddeev, V. Fock, A. Gerasimov, N.
Nekrasov, A.  Rosly and O. Tirkkonen for discussions.

\vskip 1.0cm

\baselineskip 0.55cm

{\bf References}

\vskip 0.5cm

\begin{enumerate}

\item L.D. Faddeev and L.A. Takhtajan, {\it Hamiltonian Methods
in the
Theory of Solitons} (Springer Verlag 1987)

\item A. Das, {\it Integrable Models} (World Scientific 1989)

\item J.J. Duistermaat and G.J. Heckman, Inv. Math. {\bf 69}
(1982)
259; and {\it ibid} {\bf 72} (1983) 153

\item N. Berline and M. Vergne, Duke Math. Journ. {\bf 50}
(1983) 539

\item M.F. Atiyah and R. Bott, Topology {\bf 23} (1984) 1

\item J.-M. Bismut, Comm. Math. Phys. {\bf 98} (1985) 213; and
{\it
ibid.} {\bf 103} (1986) 127; and J. Funct. Anal. {\bf 62}
(1985) 435

\item N. Berline, E. Getzler and M. Vergne, {\it Heat Kernels
and
Dirac Operators} (Springer verlag, Berlin, 1991)

\item  M.F. Atiyah, Asterisque {\bf 131} (1985) 43

\item E. Witten,  J. Geom. Phys. 9 (1992) 303

\item M. Blau, E. Keski-Vakkuri and A.J. Niemi, Phys. Lett.
{\bf B246} (1990) 92;

\item A. Hietam\"aki, A.Yu. Morozov, A.J. Niemi and K. Palo,
Phys.
Lett. {\bf 263B} (1991) 417; A. Yu. Morozov, A.J. Niemi and
K.  Palo, Phys.  Lett. {\bf 271} (1991) 365; and Nucl. Phys.
{\bf B377} (1992) 295

\item A.J. Niemi and O. Tirkkonen, Phys. Lett. {\bf B293} (1992)
339

\item A.J. Niemi and O. Tirkkonen, preprint hep-th-9301059

\end{enumerate}

\end{document}